\documentclass[usenatbib]{basi}

\usepackage[T1]{fontenc}
\usepackage[british]{babel}
\usepackage[varg]{txfonts}
\usepackage{rotating}
\usepackage{dcolumn}

\begin{document}
\title{Novae in $\gamma$-rays
}
\author[M. Hernanz]   
       {M. Hernanz\thanks{email: \texttt{hernanz@ieec.uab.es}}\\
       Institut de Ci\`encies de l'Espai, ICE (CSIC-IEEC)\\ Campus UAB, Facultat de Ci\`encies- C5 - parell - 2a planta\\
       E-08193 Bellaterra (Barcelona). Spain.}

\pubyear{2012}
\volume{00}
\pagerange{\pageref{firstpage}--\pageref{lastpage}}

\date{Received --- ; accepted ---}

\maketitle

\label{firstpage}

\begin{abstract}
Classical novae produce radioactive nuclei which are emitters of $\gamma$-rays in the MeV range. Some examples are the lines at 478 and 
1275 keV (from $^7$Be and $^{22}$Na) and the positron-electron annihilation emission (511 keV line and a continuum below this energy, 
with a cut-off at 20-30 keV). The analysis of $\gamma$-ray spectra and light curves is a potential unique and powerful tool both to trace the 
corresponding isotopes and to give insights on the properties of the expanding envelope determining its transparency. Another possible origin 
of $\gamma$-rays is the acceleration of particles up to very high energies, so that either neutral pions or inverse Compton processes produce 
$\gamma$-rays of energies larger than 100 MeV. MeV photons during nova explosions have not been detected yet, although several attempts 
have been made in the last decades; on the other hand, GeV photons from novae have been detected in some particular novae, in symbiotic binaries, 
where the companion is a red giant with a wind, instead of a main sequence star 
as in the cataclysmic variables hosting classical novae. Both mechanisms of $\gamma$-ray production in 
novae are reviewed, with more emphasis on the one related to radioactivities.
\end{abstract}

\begin{keywords}
   Stars: novae, cataclysmic variables, white dwarfs -- Gamma-rays -- Instrumentation for space
\end{keywords}

\section{Introduction}
The most energetic photons in the Universe, $\gamma$-rays, reveal explosive phenomena, like explosions of accreting white 
dwarfs in close binary systems as novae and/or thermonuclear (also called Type Ia) supernovae. Radioactive nuclei are synthesized,  
positrons which later annihilate with electrons are produced, and acceleration of particles in powerful shocks occur during and 
after the explosion. These processes release $\gamma$-rays, which thus are powerful tools  for the analysis of exploding white dwarfs. 
$\gamma$-rays directly trace isotopes, whereas observations at other wavelengths give only elemental 
abundances, except some measurements of CO molecular bands in the infrared, where $^{12}$CO and $^{13}$CO can be distinguished, 
thus giving the $^{13}$C/$^{12}$C ratio. Understanding the origin of the elements in the Galaxy and in the whole Universe is an important topic, 
intimately related to explosive nucleosynthesis and emission of $\gamma$-rays.

White dwarfs are the endpoints of stellar evolution of low-mass stars, with masses smaller than about 8-10~$M_\odot$. Their chemical 
composition is either a carbon-oxygen (CO) or an oxygen-neon (ONe) mixture (it can also be pure He). Typical white dwarf masses are 
$\sim 0.6~M_\odot$, with maximum value equal to 
the Chandrasekhar mass ($\sim1.4~M_\odot$). In white dwarfs there is not nuclear energy available anymore, and they 
cool down to very low luminosities (typically $L \sim10^{-4.5}~L_\odot$) when isolated, whereas they can explode when they are 
accreting matter in close binary systems.

There are two main types of binary systems where white dwarfs can accrete matter and subsequently explode. The most common case 
is a cataclysmic variable, where the companion is a main sequence star transferring hydrogen-rich matter. In this system, mass transfer 
occurs via Roche lobe overflow, 
typical orbital periods range from hours to days, and orbital separations are a few $10^{10}$~cm. As a consequence of accretion, 
hydrogen burning in degenerate conditions on top of the white dwarf leads to a thermonuclear runaway and a nova 
explosion, which does not disrupt the white dwarf (as occurs in Type Ia supernova explosions); therefore, after enough mass 
is accreted again from the companion star, a new explosion will occur. The typical recurrence time is $10^4$--$10^5$ years. In our 
Galaxy, there are about 35 classical nova explosions every year \citep{Sha97}. See papers by Starrfield and by Jos\'e in this volume for a review 
of nova models.

Another scenario where a white dwarf can explode as a nova is a symbiotic binary, where the white dwarf accretes matter from the stellar 
wind of a red giant companion. Typical orbital periods for these systems are a few 100 days and orbital separations are 
$10^{13}$--$10^{14}$~cm, both larger than in cataclysmic variables. This scenario leads to more frequent nova explosions than in 
cataclysmic variables, with typical recurrence periods smaller than 100 years. There are about ten known recurrent novae, for which 
more than one outburst has been recorded; about four of them are in binaries with a red giant companion \citep{Scha10}.
It is worth mentioning that recurrent novae are indeed interesting objects, since the mass of the white dwarf is expected to increase after 
each eruption - at least in some cases - and thereof they can explode as Type Ia supernovae, when the white dwarf reaches the 
Chandrasekhar mass. In most  classical novae, on the contrary, it is not expected that the mass of the white dwarf increases after 
successive outbursts, because some white dwarf core material is dredged-up and ejected as a consequence of the explosion. A hot topic 
of astrophysics is to determine the scenario where Type Ia supernovae - used as standard candles for cosmological purposes - explode; 
although some observational advances have been made recently, this question is not yet answered. It 
is thus of the maximum interest to understand recurrent nova explosions and their differences with respect to classical novae; very 
high energy (VHE) $\gamma$-rays can help for such a purpose since they give insights on the whole mass ejection process. 

The potential role of novae as $\gamma$-ray emitters was already pointed out in the 70's \citep{CH74}. 
Clayton and Hoyle stated that observable $\gamma$-rays 
from novae would come from electron-positron annihilation, with positrons 
from $^{13}$N, $^{14}$O, $^{15}$O and $^{22}$Na decays, as well as 
a result of the decay of $^{14}$O and $^{22}$Na to 
excited states of $^{14}$N and $^{22}$Ne nuclei, which de-excite by emitting photons of 2.312 and 1.274 MeV 
respectively. Seven years later, Clayton \citep{Cla81} noticed that another $\gamma$-ray line 
could be expected from novae, when $^{7}$Be transforms (through an electron capture) to an excited 
state of $^{7}$Li, which de-excites by emitting a photon 
of 478 keV. The original idea came from \cite{AR81} and both works were 
inspired by the pioneering papers mentioning the possibility of $^{7}$Li 
synthesis in novae \citep{AN75,Sta78}. In fact, $^{7}$Li production in novae 
was and is still a crucial topic \citep{Her96,Rom99}, since galactic $^{7}$Li is not well 
accounted for by other sources, either stellar (AGB stars), interstellar 
(spallation reactions by cosmic rays) or cosmological (Big Bang).

Except for the role of $^{14}$O, which is now known to be minimal, the main ideas presented in the above mentioned seminal papers 
have remained unchanged. These pioneering papers made predictions of detectability, both for the contemporaneous short mission 
HEAO 3 (High Energy Astrophysics Observatory, 1979-1981) and for the future CGRO 
(Compton Gamma-Ray Observatory, 1991-2000), already planned by NASA at that epoch, which were quite 
optimistic. Unfortunately, later computations with updated input physics (especially the nuclear reaction rates) provided lower yields and 
consequently less optimistic expectations (see recent reviews by \cite{Her02,Her08} and previous ones by 
\cite{Lei91,Lei93,Lei97}). In fact, an unfortunate combination of low emission fluxes and low sensitivities of $\gamma$-ray instruments 
makes detection of novae really difficult, even with the current instruments on board the INTEGRAL satellite (INTErnational Gamma-RAy 
Laboratory), launched 10 years ago, on October 17, 2002. 

The detection of $\gamma-$ray photons in the MeV range is extremely challenging; up to 
now, radioactivities from novae have not been detected. But interestingly 
enough, there has been at least one detection in VHE $\gamma$-rays, by the Fermi satellite \citep{Abdo10}, revealing the 
powerful shocks between the nova ejecta and the red giant wind of the companion, in the particular case of a recurrent nova 
in a symbiotic binary, 
V407 Cyg. A second detection of a nova - Nova Sco 2012 - with the Fermi Large Area Telescope (LAT) has been announced 
very recently (ATel \#4284, 28 July 2012, Cheung et al.).

In this paper we review the main mechanisms of $\gamma$-ray production in novae, both those related to radioactivities and to particle acceleration in shocks. 
Models of $\gamma$-ray emission by CO and ONe novae follow, as well as the results of observational campaigns with past and current satellites. Finally, 
planned instruments for future $\gamma$-ray missions are briefly discussed.

\section{Why do novae emit $\gamma$-rays?}

\subsection{Radioactivities in novae}

Novae can emit $\gamma$-rays because some of the radioactive nuclei they 
synthesize during the hydrogen thermonuclear runaway either experiment electron 
captures or are $\beta^+$-unstable (thus emitting 
positrons), decaying in some cases to nuclei in excited states which
de-excite to their ground states by emitting photons at particular energies in the 
$\gamma$-ray range. The main $\beta^+$-unstable nuclei produced by the carbon-nitrogen-oxygen 
(CNO) cycle out of equilibrium are $^{13}$N, $^{14,15}$O, $^{17,18}$F. The positrons themselves annihilate with electrons and 
therefore also produce $\gamma$-ray emission: 511 keV line plus a 
continuum below this energy \citep{LC87,Gom98}.
The other relevant radioactive nuclei are synthesized through proton-proton chains, 
$^{7}$Be, or through the NeNa or MgAl ``cycles'', $^{22}$Na and $^{26}$Al; they decay 
(or, for $^{7}$Be, suffer an electron capture) to excited states of daughter 
nuclei (i.e. $^{7}$Li, $^{22}$Ne and $^{26}$Mg), which de-excite to their ground 
state emitting $\gamma$-ray photons at particular energies (see Table \ref{tab:radioactivities} for details). 
Notice that there are some differences between carbon-oxygen (CO) and 
oxygen-neon (ONe) novae: $^{7}$Be synthesis is favored in the former ones whereas $^{22}$Na and $^{26}$Al 
are mainly produced in ONe novae. 

The lifetimes of radioactive nuclei should not be shorter than the timescale of envelope transparency, in order to allow the 
$\gamma$-rays to escape; this excludes the very short-lived $^{14}$O, $^{15}$O and 
$^{17}$F (lifetimes 102, 176 and 93 s respectively).  $\gamma$-rays give thus a direct insight both on the 
nucleosynthesis during the nova explosion, and on the global properties of the expanding envelope, mainly dependent on density, 
temperature, chemical composition and velocity profiles, which determine its transparency 
to $\gamma$-rays. Comptonization and photoelectric absorption take place in nova envelopes and play a crucial role for the shape 
of the emitted spectra (see details in \cite{Gom98}). The emitted $\gamma$-rays can be potentially detected, either in individual 
objects or also as a diffuse emission from the cumulative $\gamma$-ray output of several objects in the galaxy, whenever the lifetime of a 
given isotope is longer than the average period between two successive events producing it. This last possibility applies to the case 
of $^{26}$Al, which we don't include in this review, because novae are not the major contributors to the Galactic content of 
this isotope and also it can't be detected in individual novae (see \cite{Her02,Her08}). The other potential case of cumulative emission 
is $^{22}$Na, but detectability is also quite difficult (see \cite{Jea00} for details). 


\begin{table}[h]
\caption{Radioactivities in nova ejecta.}
\label{tab:radioactivities}
\begin{tabular}{ccccc} 
\hline \hline
Isotope   & Lifetime       & Main disintegration  & Type of emission
          & Nova type\\
          &                            &  process                      & 
          &\\
\hline
$^{13}$N  & 862~s          & $\beta^+$-decay              
                                             & 511~keV line and
          & CO and ONe\\      
          &                &         &    continuum        & \\
$^{18}$F  & 158~min        &  $\beta^+$-decay      
                                              & 511~keV line and 
          & CO and ONe\\
          &                &         &   continuum         & \\
$^{7}$Be  & 77~days        & $e^-$-capture      
                                              & 478~keV line 
          & CO\\
$^{22}$Na & 3.75~years  &  $\beta^+$-decay
                                              & 1275 and 511~keV lines  
          & ONe\\
$^{26}$Al & 10$^{6}$~years &  $\beta^+$-decay
                                              & 1809 and 511~keV lines 
          & ONe\\
\hline \hline
\end{tabular}
\end{table} 


\subsection{Particle acceleration and VHE $\gamma$-rays}

A second mechanism of $\gamma$-ray production in novae is related to acceleration of particles in shock waves, 
expected to be relevant only when the nova ejecta collides with the wind of the red giant companion, 
i.e., only in the case of recurrent novae in symbiotic binaries. A good example is the recurrent nova RS Oph, which had its two last 
eruptions in 1985 and 2006. Once the nova explodes, an expanding shock wave 
forms which sweeps the red giant wind: the system behaves as a ``miniature" supernova remnant, evolving much faster and being 
much dimmer. It was predicted that this nova could accelerate cosmic rays \citep{TH07,HT12}, with the ensuing emission of  
$\gamma$-rays with energies larger than 100 MeV coming mainly from neutral pion, $\pi^0$, production. Such emission from RS Oph 
would have been 
detected by the Fermi satellite, but it was not in orbit yet in 2006. A more recent object, V407 Cyg (another nova occurring in a symbiotic binary 
with a red giant companion) has been detected by Fermi \citep{Abdo10}, thus confirming our previous theoretical predictions for RS Oph.

\section{Models}

\subsection{Emission from radioactivities in novae}

The $\gamma$-ray signatures of classical novae mainly depend on their yields of 
radioactive nuclei. CO and ONe novae differ 
in their production of $^{7}$Be and $^{22}$Na, whereas they 
synthesize similar amounts of $^{13}$N and $^{18}$F. Therefore, CO novae should 
display line emission at 478 keV related to $^{7}$Be decay, whereas for ONe 
novae line emission at 1275 keV related to $^{22}$Na decay is expected. In 
both nova types, there should be as well line emission at 511 keV related to 
e$^-$--e$^+$ annihilation, and a continuum produced by Comptonized 511 keV 
emission and positronium decay (see Table \ref{tab:radioactivities}).

The shape an intensity of the $\gamma$-ray output of novae, related to the radioactive decay of 
the unstable isotopes synthesized during the explosion, as well as its temporal 
evolution does not depend only on the amount of $\gamma$-ray photons produced, 
but also on how they propagate through the expanding envelope and ejecta \citep{LC87,Gom98}. 
Several interaction processes affect the propagation of photons, i.e. Compton 
scattering, e$^-$--e$^+$ pairs production and photoelectric absorption. 

The treatment of positron annihilation deserves particular attention. When a 
positron is emitted, it can either escape without interacting with the nova 
expanding envelope or annihilate with an ambient electron. It can be safely 
assumed that in nova envelopes positrons thermalize before annihilating. This 
approximation is wrong in less than 1\% of cases in an electronic plasma, 
according to \cite{LC87}. Once thermalized, the positron covers a negligible distance and 
then annihilates. For densities and temperatures typical of nova envelopes, positrons form 
positronium (positron-electron system) in $\sim 90$\% of annihilations 
\citep{LC87}, while in the remaining 10\% of cases they annihilate directly. 
Positronium is formed in singlet state 25\% of the time, leading to the emission 
of two 511 keV photons, and in triplet state 75\% of the time, leading to a 
three-photon annihilation. The continuum spectrum of photons produced by 
triplet state was obtained by \cite{OP49}. In summary, once a positron is 
produced, its trip should be followed until it escapes or covers the average 
energy-loss distance. In the latter case it produces positronium 90\% of the 
time, leading to triplet to singlet annihilations in 3:1 proportion, while in 
10\% of the cases it annihilates directly.

\begin{figure}
\centerline{
\hspace{2cm}
\includegraphics[width=8.5cm]{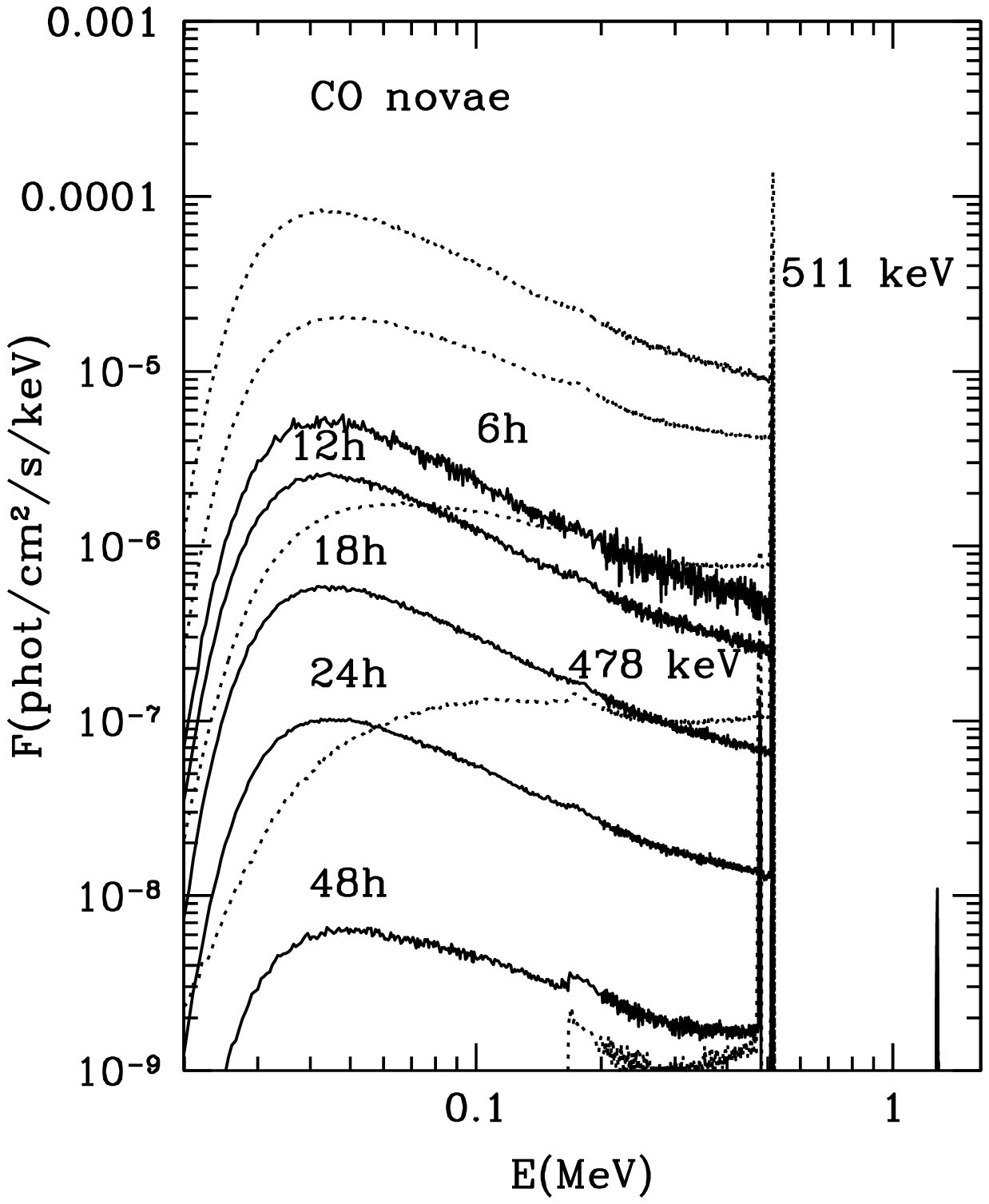}
\hspace{-3cm}
\includegraphics[width=8.5cm]{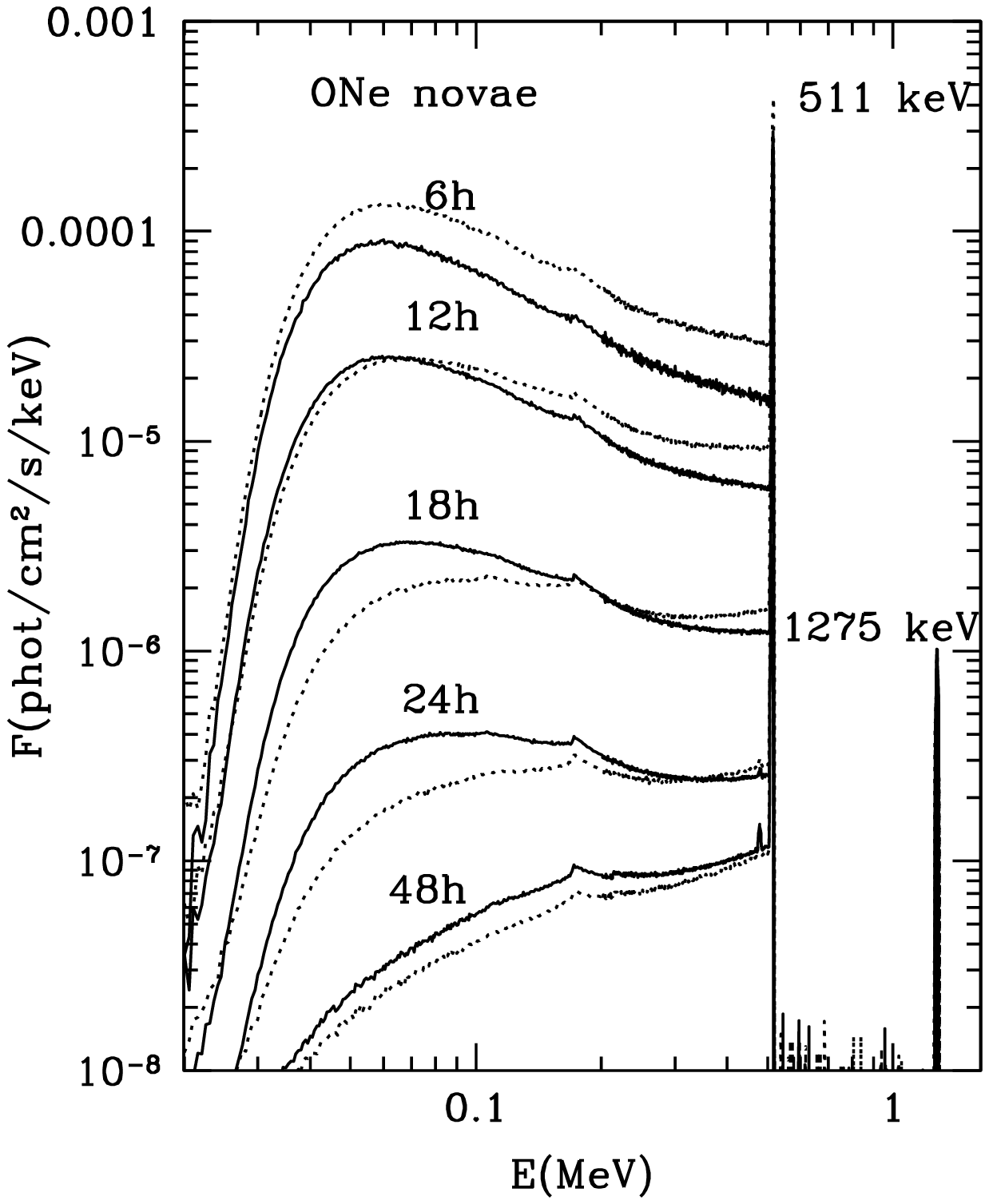}
}
\caption{Left panel: Spectra of CO novae of masses 0.8 (solid) and 1.15 M$_\odot$ (dotted) at different epochs after T$_{peak}$; 
Right panel: Same for ONe novae of masses 1.15 (solid) and 1.25 M$_\odot$ (dotted).}
\label{fig:specCO_ONe}
\end{figure}

A Monte Carlo code, based on the method described by \cite{PSS83} and \cite{AS88}, 
was developed by \cite{Gom98} to compute the $\gamma$-ray output of novae. 
The temporal evolution of the whole $\gamma$-ray spectrum of four  
representative models is shown in 
Figure \ref{fig:specCO_ONe}. 
The most prominent features of the spectra are the annihilation line at 511 keV and 
the continuum at energies between 20-30 keV and 511 keV (in both nova 
types), the $^{7}$Be line at 478 keV in CO novae, and the $^{22}$Na line 
at 1275 keV in ONe novae. A few hours after the outburst, 
when transparency increases, the back-scattering of the 511 keV photons produces a feature at 170 keV. 
The main difference between spectra of CO and 
ONe novae are, as expected, the long-lived lines: 478 keV in CO novae as compared with 
1275 keV in ONe novae, which directly reflect the different 
chemical composition of the expanding envelope ($^{7}$Be-rich in CO novae and 
$^{22}$Na-rich in ONe ones). 

\subsubsection{511 keV line and continuum emission from electron-positron annihilation}

\begin{figure}
\centerline{
\hspace{2cm}
\includegraphics[width=8.5cm]{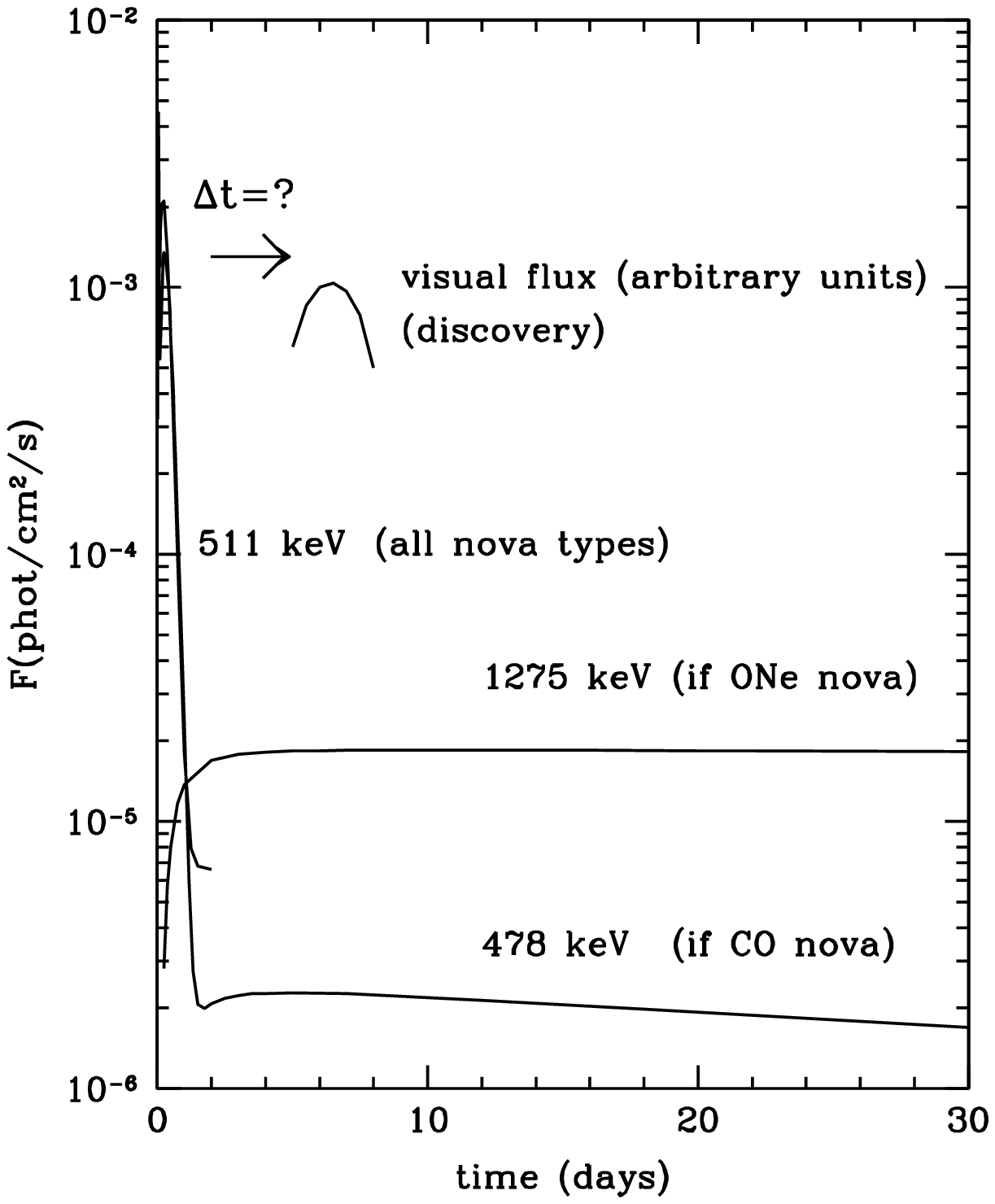}
\hspace{-3cm}
\includegraphics[width=8.5cm]{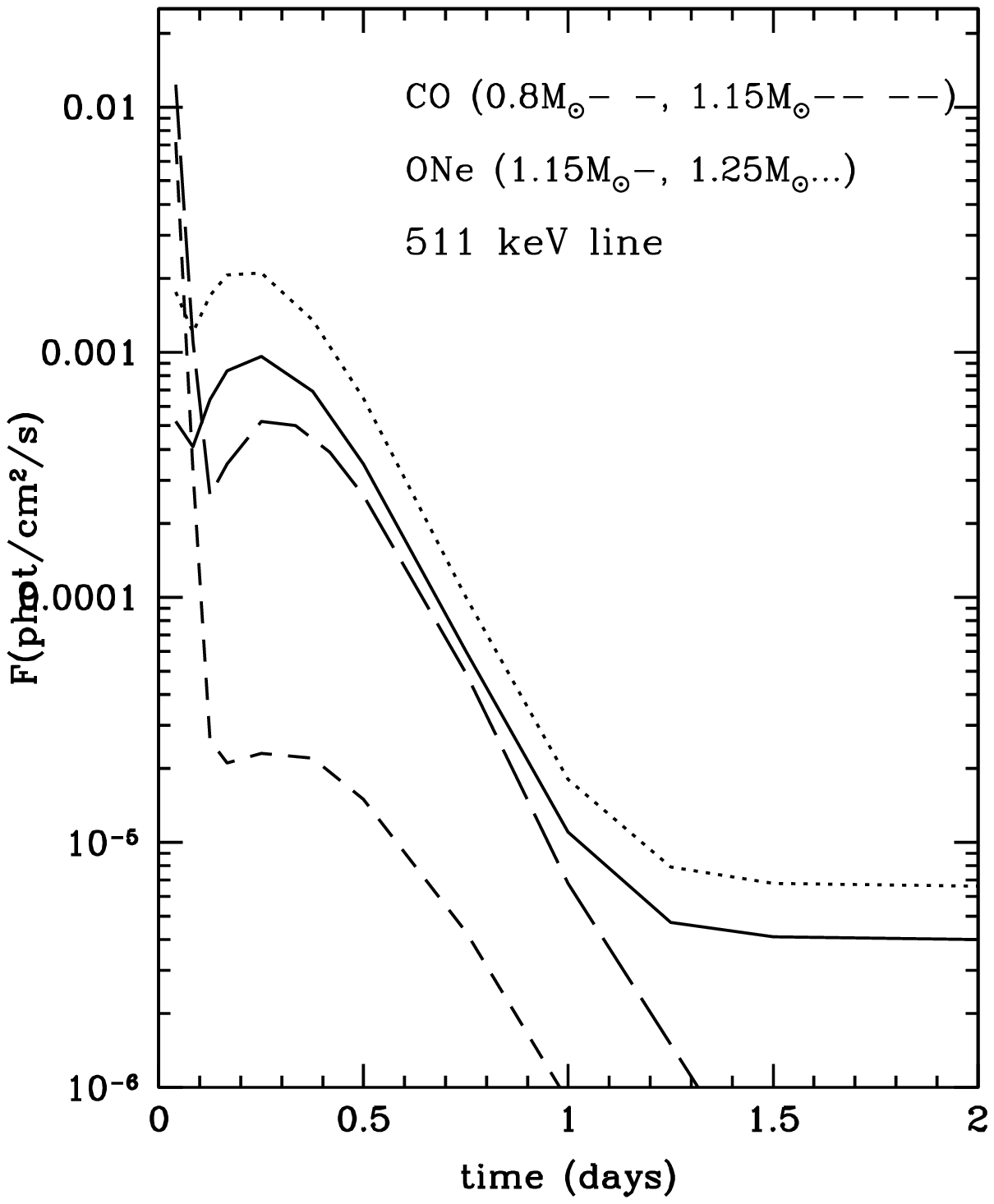}
}
\caption{Left panel: Visual light curve (in arbitrary units, shifted vertically for illustrative purposes) compared to $\gamma$-ray light curves 
for the lines at 511, 478 and 1275 keV. Notice that the maximum in visual occurs after the maxima in $\gamma$-rays. 
Right panel: Light curves of the 511 keV line for CO and ONe 
novae \citep{Gom98}.}
\label{fig:lcall_511}
\end{figure}

\begin{figure}
\centerline{
\hspace{2cm}
\includegraphics[width=8.5cm]{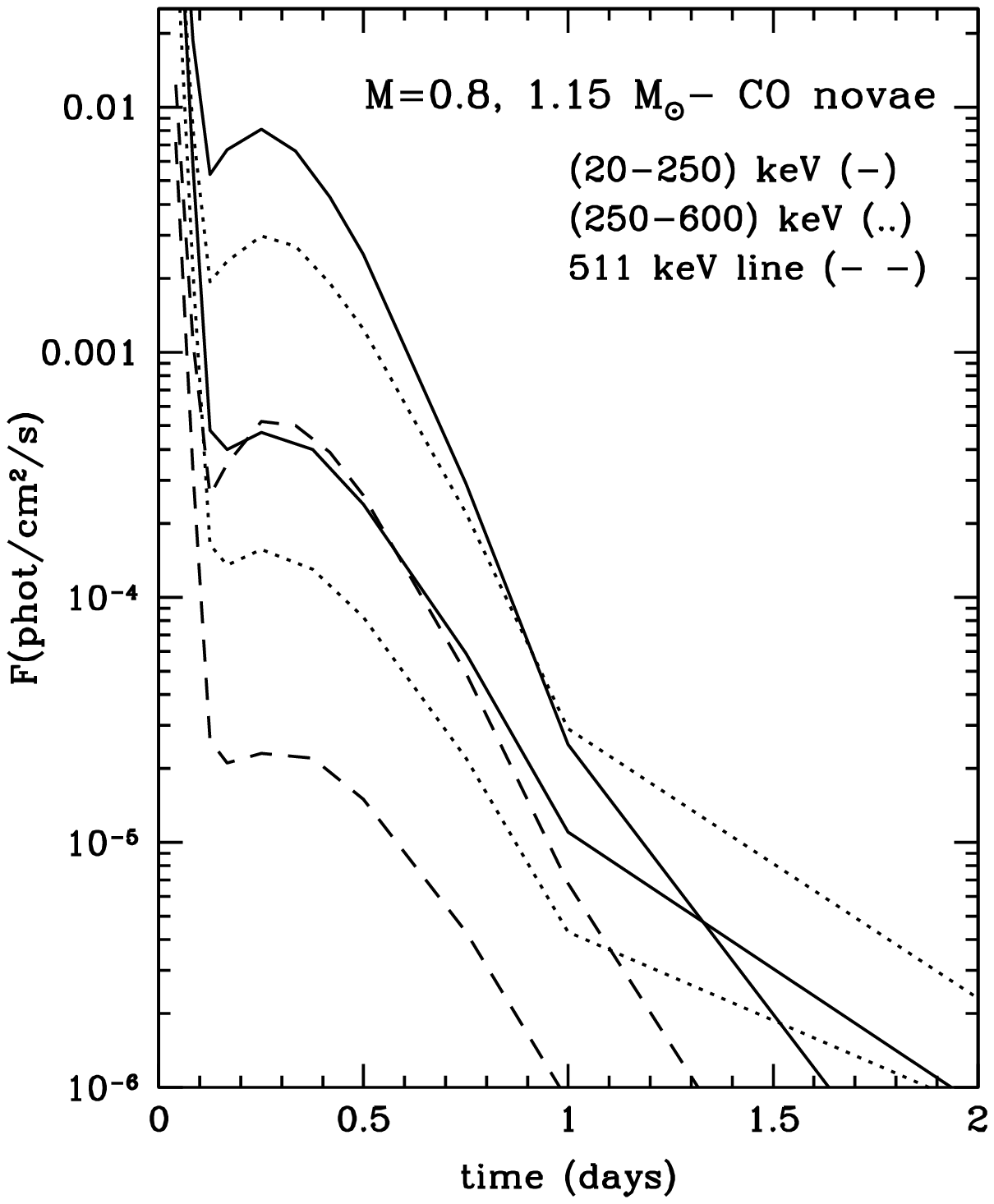}
\hspace{-3cm}
\includegraphics[width=8.5cm]{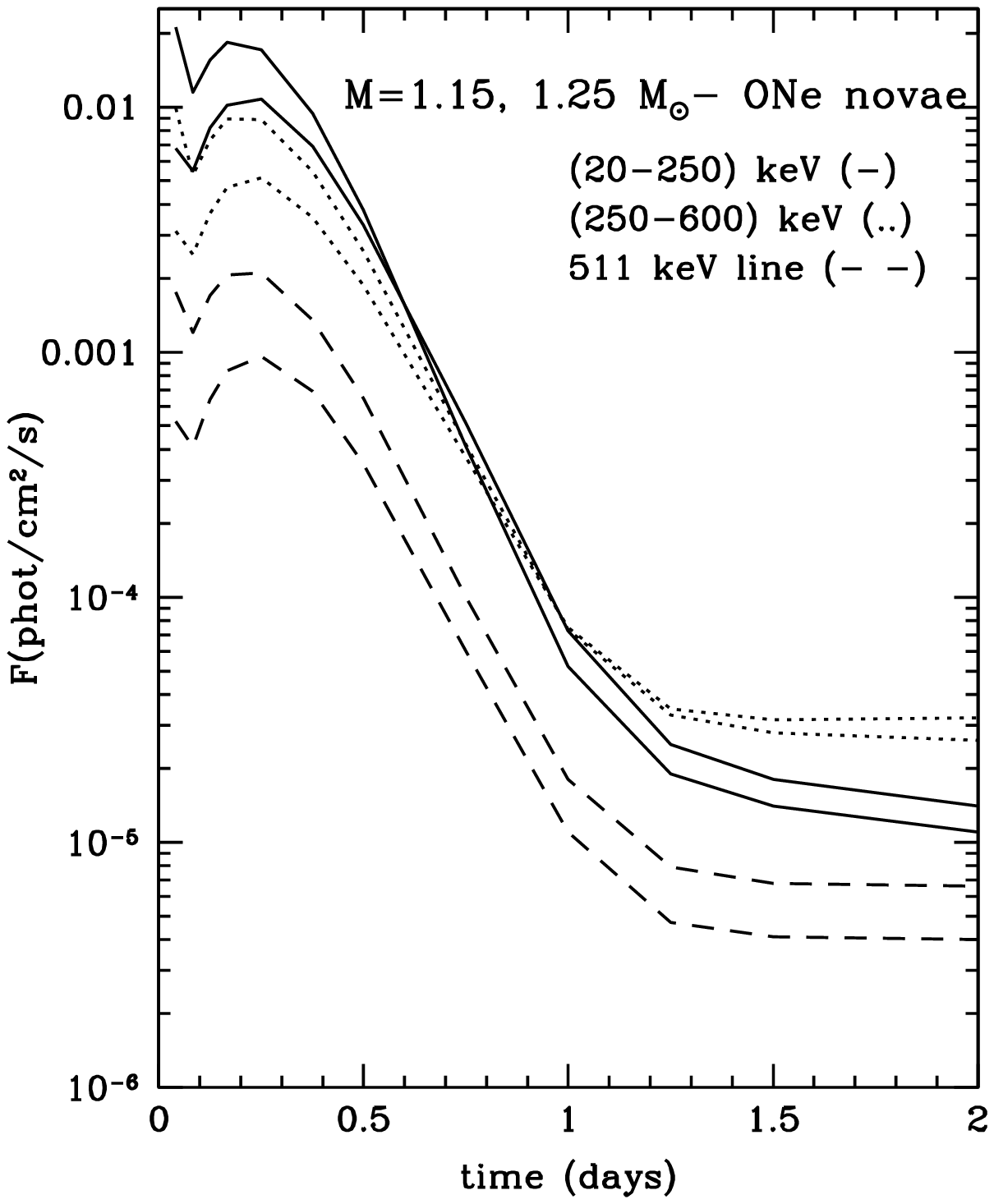}
}
\caption{Light curves of two continuum bands below 511 keV for CO (left) and ONe (right)
novae. The light curve of the 511 keV line is also shown for comparison. The upper 
curves correspond to the more massive novae.  But at later epochs, for ONe novae, 
the most massive nova emits a slightly smaller flux, except for the 511 keV line, 
because of larger transparency.}
\label{fig:lccontCO_ONe}
\end{figure}

The early $\gamma$-ray emission, or prompt emission, of novae is related 
to the disintegration of the very short-lived $\beta^+$-unstable 
isotopes $^{13}$N and $^{18}$F. The radiation is emitted as 
a line at 511 keV (direct annihilation of positrons and singlet state 
positronium), plus a continuum. The continuum is related with both the 
triplet state positronium continuum and the Comptonization of the photons 
emitted in the line. 
There is a sharp cut-off at energies 20-30 keV (the 
exact value depending on the envelope composition) because of photoelectric 
absorption. A better insight into the early $\gamma$-ray emission is obtained 
from inspection of the light curves shown in 
Figures \ref{fig:lcall_511} (right panel) and \ref{fig:lccontCO_ONe}.
The largest flux is emitted in the 
(20-250) keV range, since the continuum has its maximum at $\sim$60 keV (ONe 
novae) and at $\sim$45 keV (CO novae), followed by the flux in the 
(250-511) keV range (excluding the 511 keV line) and the flux in the 
511 keV line. The two maxima in the light curves of the 511 keV line correspond 
to $^{13}$N and $^{18}$F decays, but the first maximum is difficult to resolve 
because its duration is really short; in addition, it is very model dependent: 
only $^{13}$N in the outermost zones of the envelope can be seen in $\gamma$-rays 
because of limited transparency at very early epochs and, therefore, 
the intensity of the first maximum depends on the efficiency of convection. 
This first maximum gives thus an important insight into the dynamics of the envelope after 
peak temperature is attained at its base.

The annihilation emission is the most intense $\gamma$-ray feature expected from novae, 
but unfortunately it has a very short duration, because of the short lifetime 
of the main positron producers ($^{13}$N and $^{18}$F). There are also positrons 
available from $^{22}$Na decay in ONe novae, but these contribute much less 
(they are responsible for the {\it plateau} at a low level, between $10^{-6}$ and $10^{-5}$ 
phot cm$^{-2}$ s$^{-1}$ for d=1kpc; see 
Figure \ref{fig:lcall_511}, right panel). 
These positrons do not contribute all the time, because after roughly one week the envelope is so transparent 
that $^{22}$Na positrons escape freely without annihilating. In summary, annihilation 
radiation lasts only $\sim 1$ day at a high level, and one to two weeks at a lower level 
{\it plateau} (the latter only in ONe novae).

Another important fact is that annihilation radiation is emitted well before the 
visual maximum of the nova, i.e. before the nova is discovered optically 
(see Figure \ref{fig:lcall_511}, left panel).
This early appearance of $\gamma$-rays from electron-positron annihilation makes their detection 
through pointed observations almost impossible. Only wide field of view instruments, 
monitoring continuously the sky in the appropriate energy range can detect it.

\subsubsection{Line emission at 478 and 1275 keV from $^{7}$Be and $^{22}$Na}

\begin{figure}
\centerline{
\hspace{2cm}
\includegraphics[width=8.5cm]{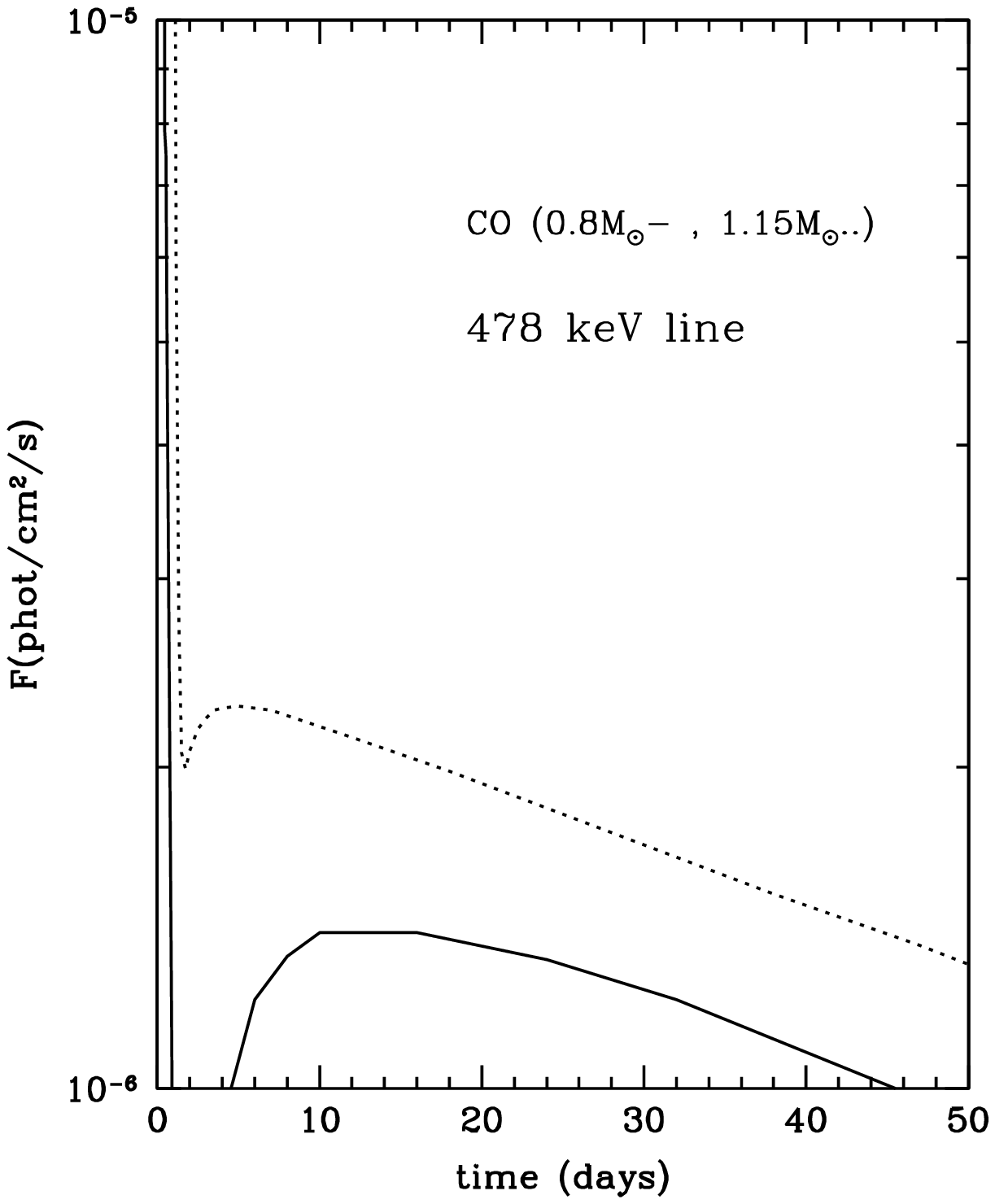}
\hspace{-3cm}
\includegraphics[width=8.5cm]{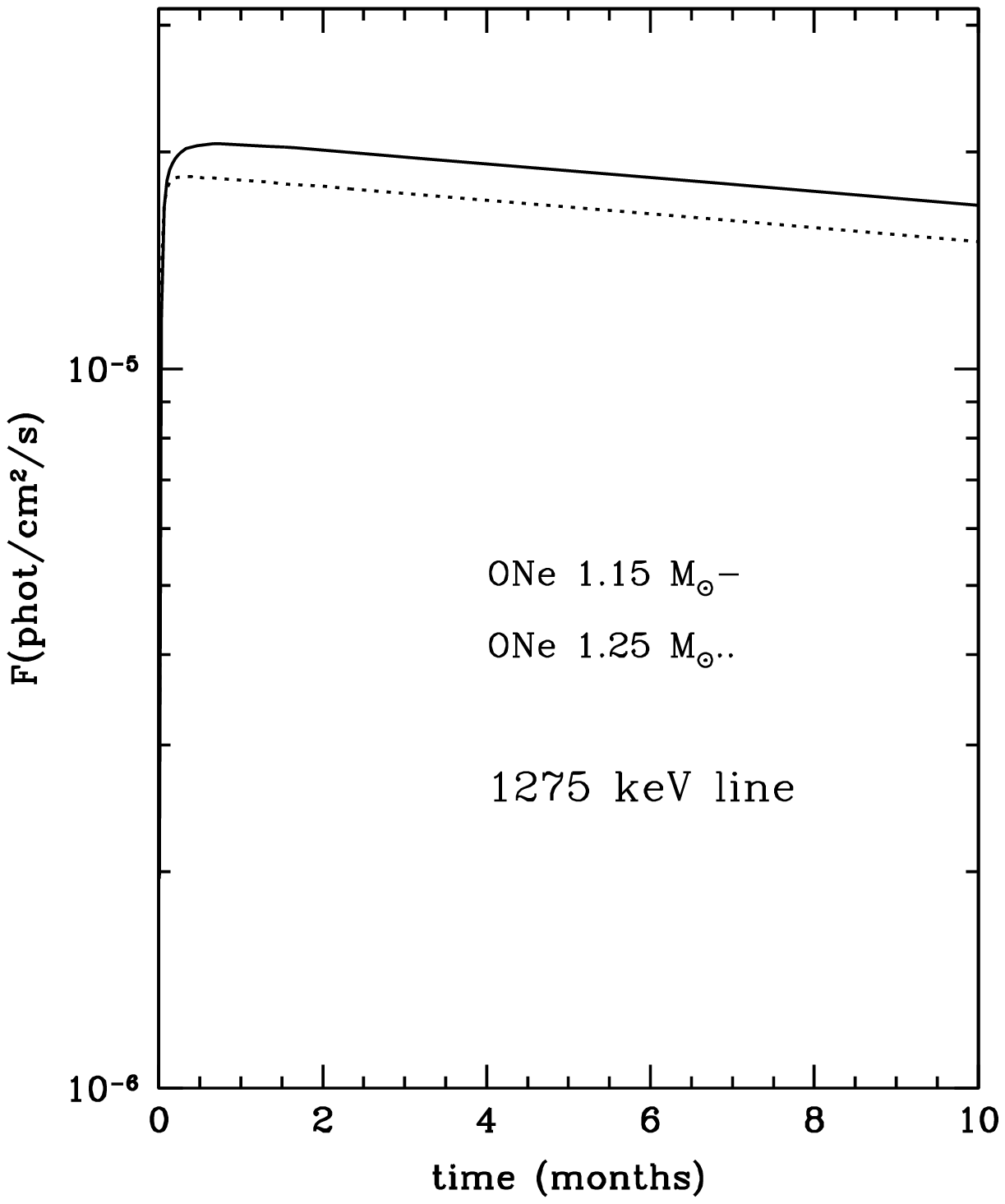}
}
\caption{Left panel: Light curves of the 478 keV line for the two CO nova models of Figure \ref{fig:specCO_ONe}. 
Right panel: Same for ONe novae models.}
\label{fig:lcBe_Na}
\end{figure}

Line emission at 478 keV, related to de-excitation of the $^{7}$Li which results from 
electron captures on $^{7}$Be, is the most 
distinctive feature in the $\gamma$-ray spectra of CO novae (besides the 
annihilation line and continuum, common to all types of novae), as shown in 
Figure \ref{fig:specCO_ONe}. 
The reason is that these novae are more prolific producers of $^{7}$Be 
than ONe novae.  
The light curves of the 478 keV line are shown in 
Figure \ref{fig:lcBe_Na}:  
the flux reaches its maximum at day 13 and 5 in the more and less opaque 
models, with total masses 0.8 and 1.15 M$_\odot$, respectively. The width of the line 
is 3 and 8 keV for the 0.8 and 1.15 M$_\odot$ CO novae, respectively.
The maximum flux 
is around $10^{-6}$phot cm$^{-2}$ s$^{-1}$, for d=1kpc. There is a 
previous maximum, which has nothing to do with the envelope's content of $^{7}$Be, 
but with the strong continuum related to the annihilation of 
$^{13}$N and $^{18}$F positrons.

The $^{22}$Na line at 1275 keV appears only in ONe novae, because CO novae do not 
synthesize this isotope. The rise phase of the 1275 keV line 
light curves 
(see Figure \ref{fig:lcBe_Na}) 
lasts between 10 (1.25 M$_\odot$) and 20 days (1.15 M$_\odot$). 
Soon after the maximum, the 
line reaches the stable decline phase dictated by the lifetime of  $^{22}$Na, 
3.75 years; during this phase, the line intensities directly reflect the amount 
of $^{22}$Na ejected mass. The corresponding fluxes at maximum are around 
$10^{-5}$phot cm$^{-2}$ s$^{-1}$ at d=1kpc. 
The width of the line is around 20 keV, which is a handicap for its  
detectability with instruments having good spectral resolution, which are best suited for narrow lines 
(e.g., Ge detectors of SPI on board INTEGRAL).

\subsection{Emission from accelerated particles}

The blast wave evolution of RS Oph during its 2006 outburst is shown in Figure 1 of \cite{TH07}, where the time dependence of the forward shock velocity 
as deduced from IR spectroscopic observations is compared to that from the X-ray observations with RXTE \citep{Sok06} and Swift \citep{Bod06}. 
The usual relation for a test-particle 
strong shock $v_s = [(16/3) (k T_s)/(\mu m_{\rm H})]^{1/2}$ was used to get shock velocities from X-ray fits of the post-shock temperature $T_s$ 
($k$ is the Boltzmann constant and $\mu m_{\rm H}$ the mean particle mass). $v_s$ deduced from X-ray observations appear to be significantly smaller 
than those from IR data. Therefore the question was: what made the X-ray measurements of $v_s$ smaller than those 
from IR data? Another caveat was to understand why the cooling phase started as early as 6 days (see Figure 1 of \cite{TH07}), 
when $T_s$ was about $10^8$~K and thus radiative cooling was not important. Acceleration of cosmic rays in RS Oph can answer both questions.

Acceleration of cosmic rays influences the evolution of a nova remnant, mainly because the escape of the highest energy 
particles from the blast wave region leads to energy loss. Then, a good agreement between the IR and X-ray measurements of the shock 
temperature in RS Oph was obtained 
(see Figure 3a in \cite {TH07}), with a moderate acceleration efficiency, $\eta_{\rm inj} \sim 10^{-4}$; this parameter is the fraction of total shocked protons 
in protons injected from the thermal pool into the diffusive acceleration process. The fraction of the total energy flux processed by the shock that escaped 
via diffusion away from the shock system of the highest energy particles is shown in Figure 3b from \cite {TH07}. The corresponding energy loss rate was 
estimated to be $\sim 2 \times 10^{38} (t/{\rm 6~days})$~erg~s$^{-1}$, which is approximately 100 times the bolometric luminosity of the post-shock plasma 6 days after outburst. This means that energy loss via 
escape of accelerated particles is much more efficient than radiative losses to cool the shock.

A prediction of the $\gamma$-ray emission associated to the accelerated particles could be made. The neutral pion ($\pi^0$) production was calculated from 
the gas density in the red giant wind (i.e. $(dM/dt)_{\rm wind}$ and $v_{\rm wind}$) and the cosmic-ray energy density required to explain the IR and X-ray 
observations. The Inverse Compton (IC) contribution was estimated from the nonthermal synchrotron luminosity $L_{\rm syn}$ and the ejecta luminosity 
$L_{\rm ej}$: $L_{\rm IC}=L_{\rm syn} \times U_{\rm rad}/(B^2/8 \pi)$, where $U_{\rm rad} \sim L_{\rm ej}/(4\pi cR_s^2)$ ($R_s$ is the shock radius and $c$ 
the speed of light) and $B$ is the post-shock magnetic field. According to early radio detections of RS Oph at frequencies lower than 1.4 GHz by 
\cite{Kanth07}: $L_{\rm syn} \sim 5 \times 10^{33} (t/{\rm 1~day})^{-1.3}$~erg~s$^{-1}$. The ejecta luminosity was assumed to be equal to the 
Eddington luminosity (i.e. approximately $10^{38}$~erg~s$^{-1}$). We then found $L_{IC} \sim L_{\rm syn}$ and could conclude 
that $\gamma$-rays come mainly from $\pi^0$ production (see Figure 3 in \cite{HT12}).

Finally, predicted light curves for $\gamma$-ray energies $E_\gamma > 100$~MeV and $E_\gamma > 30$~GeV were compared in Figure 4 from \cite{HT12} 
to the Fermi/LAT sensitivities for 1 week and 1 month observation times. The conclusion was that RS Oph (2006) would have been detected by 
Fermi/LAT. Since the origin of the high energy $\gamma$-rays is the collision of the nova blast wave with the dense wind of the red giant companion, such a situation 
can only be found in novae occurring in symbiotic binaries. 

We have made a preliminary estimate of the $\gamma$-ray flux from neutral pion decay in V407 Cyg, getting 
$F_\gamma(E_\gamma > {\rm 100~MeV}) \sim 10^{-6}$~photons~cm$^{-2}$~s$^{-1}$, for an estimated mean post-shock density 
for a few days after outburst of $\sim 2 \times 10^9$~cm$^{-3}$, a nova energy output of $\sim 10^{44}$~erg and a distance of 2.7 kpc. 
This result is consistent with the peak $\gamma$-ray flux detected by Fermi/LAT about 3--4 days after the optical outburst \citep{Abdo10}.

\section{Observations}
\subsection{$^{7}$Be and $^{22}$Na lines}

First observations of $\gamma$-rays with possible origin in novae were performed with 
a balloon-borne $\gamma$-ray telescope using a large volume lithium-drifted-germanium 
[Ge(Li)] crystal as detector \citep{LCW77}. During the second flight of 
the instrument, in 1976, two novae were pointed, Nova Cygni 1975 and Nova Serpentis 1970, to 
search for their 1275 keV emission. Only $2\sigma$ flux upper limits around 
$10^{-3}$~phot cm$^{-2}$ s$^{-1}$ were obtained.

The $\gamma$-ray spectrometer (GRS) 
on board the Solar Maximum Mission (SMM),  made another attempt to detect 
the $^{22}$Na line from novae. In data accumulated from 1980 to 1987, only upper limits 
on the 1275 keV emission by individual neon novae (Nova Cyg 1975, Nova Cr A 1981, Nova Aql 1982, 
Nova Vul 1984\#2) and on the whole galactic center were obtained, between 
1.2 and 25 $\times 10^{-4}$~phot cm$^{-2}$ s$^{-1}$ \citep{Lei88}.
The Oriented Scintillation Spectrometer (OSSE) on board the Compton Gamma-Ray Observatory 
(CGRO) also searched for the 1275 keV line \citep{Har96}; the $3 \sigma$ upper limit they found for the 
central radian of the Galaxy was $2.25 \times 10^{-4}$ phot cm$^{-2}$ s$^{-1}$ rad$^{-1}$, worse than 
the limit obtained with SMM. 

The most extensive and recent observations of the $^{22}$Na line at 1275 keV from novae 
has been performed by \cite{Iyu95}, with the COMPTon TELescope (COMPTEL) instrument onboard the 
CGRO. COMPTEL observed several novae during 
the period 1991-1993, five of them of the neon type , i.e., those expected to be the main producers of $^{22}$Na: 
Nova Her 1991, Nova Sgr 1991, Nova Sct 1991, Nova Pup 1991 and 
Nova Cyg 1992. None was 
detected. The average $2\sigma$ upper limit for any nova of the ONe type in the 
galactic disk was around $3\times 10^{-5}$ phot cm$^{-2}$ s$^{-1}$, which translated into an 
upper limit of the ejected $^{22}$Na mass around $3.7\times 10^{-8}$ M$_\odot$, for the adopted 
distances to the observed novae. This limit was constraining for the currently available 
models at the epoch \citep{Sta92,Sta93,Pol95}, but not for the more recent models \citep{JH98}. 
The main reason for the discrepancy between models of different groups 
(\cite{JH98} versus \cite{Pol95} and \cite{Sta98}) is that old models were based on the explosion on 
ONeMg white dwarfs, whereas recent models adopt ONe white dwarfs as underlying cores, because more recent 
evolutionary calculations of stellar evolution predict much lower magnesium 
abundances \citep{Rit96,Dom93}, than the parametrized calculations of hydrostatic carbon burning nucleosynthesis by \cite{AT69} 
previously used. In both model series, some mixing between the accreted H-rich matter and the underlying 
white dwarf core is assumed. The smaller initial content of neon and magnesium makes 
$^{22}$Na synthesis much less favored. 

Regarding the $^{7}$Be line at 478 keV, the first search from the galactic center and from some 
particular novae was performed with SMM/GRS by \cite{HLS91}. Three individual novae, 
Nova Aql 1982, Nova Vul 1984 \#2 and Nova Cen 1986, were observed, and the 
upper limits derived ranged between $8.1 \times 10^{-4}$  
and $2.0 \times 10^{-3}$ ~phot cm$^{-2}$ s$^{-1}$ (see as well \cite{Har01}). The corresponding upper 
limits of $^7$Be ejected masses ranged between  $5 \times 10^{-8}$ and 
$6.3 \times 10^{-7}$ M$_\odot$ \citep{HLS91}. These fluxes and masses are well above the current theoretical 
predictions and thus, again, do not constrain the models.
 
More recent analyses have been made with the Transient 
Gamma-Ray Spectrometer (TGRS) on board the Wind satellite. 
Five novae were in the field of view of TGRS during the period 
1995-1997: Nova Cir 1995, Nova Cen 1995, 
Nova Sgr 1996, Nova Cru 1996 and Nova Sco 1997. The $3 \sigma$ 
upper limits obtained ranged from $6 \times 10^{-5}$ to 
$2 \times 10^{-4}$~phot cm$^{-2}$ s$^{-1}$, leading to ejected 
masses between $6 \times 10^{-5}$ and 
$3 \times 10^{-7}$~M$_\odot$ \citep{Har01}. 
The flux limits from TGRS were a factor of 10 better (smaller) than 
those from SMM 
observations, but the upper limits on $^7$Be ejected masses did not improve 
by the same factor, mainly because novae observed with TGRS were at larger 
distances than those observed with SMM. Upper limits on the integrated  
emission at 478 keV flux from the galactic center with SMM and TGRS were 
around $10^{-4}$~phot cm$^{-2}$ s$^{-1}$. The improvement in the limit from 
SMM to TGRS is not very noticeable, mainly because the 
equivalent aperture of TGRS (occulted region $16^{\rm o} \times 90^{\rm o}$) is smaller 
than that of SMM ($\sim 130^{\rm o}$).

\subsection{Positron-electron annihilation: 511 keV line and continuum}

The emission resulting from e$^-$-e$^+$ 
annihilation is the most intense $\gamma$-ray outcome of classical novae, but 
$\gamma$-rays are emitted 
well before the visual maximum of the nova, i.e., before the nova is discovered, 
and have a very short duration 
(see Figure \ref{fig:lcall_511}). 
Therefore, they can not be detected through 
observations pointing to a particular nova already discovered. Wide field of view 
instruments monitoring the sky in the appropriate energy range, like the old
Burst and Transient Source Experiment (BATSE) on board CGRO or TGRS on board Wind, 
or the more recent Reuven Ramaty High Energy Solar Spectroscopic Imager (RHESSI), 
and the Burst Alert Telescope (BAT) onboard Swift,
are therefore the best suited instruments for the search of the 511 keV line 
and the continuum below it. 

TGRS was very convenient to search for the 511 keV line
\citep{Har99}. Its germanium detectors had enough spectral 
resolution to separate the cosmic 511 keV line from the nova line, provided that 
the latter is a bit blue shifted (which happens only at the beginning of 
the emission phase, when material is not completely transparent yet). TGRS's 
large field of view contained five new novae during the period 1995-1997 (see 
previous section). The line at 6 hours and at 12 hours was modeled for each nova, 
according to theoretical models \citep{JH98,Gom98}. Then a comparison with background 
spectra during periods encompassing some days before the discovery date of 
each nova provided upper limits to the 511 keV line flux in 6 hours, around 
$2-3 \times 10^{-3}$~phot cm$^{-2}$ s$^{-1}$. \cite{Har99} 
deduced that their method was sensitive enough to detect novae occurring out to about 
2.5 kpc. This number has to be corrected, since it was based on the current models 
at the epoch \citep{Gom98}, which predicted too much $^{18}$F; there has been a drastic change 
in nuclear reactions affecting $^{18}$F 
synthesis \citep{Her99,Coc00}. A 
reduction of $^{18}$F synthesis in novae by 
a factor around 10 with respect to the yields adopted in \cite{Gom98} and taken 
as model templates by \cite{Har99} was found; so that, TGRS would be sensitive enough to detect 
novae at around 0.8 kpc, of any type (CO and ONe). Further reductions in $^{18}$F yields \citep{Ser03,Cha05}
make this distance even shorter.

Another instrument well suited for the detection of the prompt $\gamma$-ray 
emission from novae was BATSE on board CGRO. Before the launch of CGRO in 1991, 
\cite{Fis91} already made a prediction of the detectability of low-energy
$\gamma$-rays from novae with this instrument, based on the models of
$\gamma$-ray emission from \cite{LC87}. BATSE had the advantage of covering the whole sky 
all the time, but on the other hand it was not very sensitive and it had poor energy 
resolution. Data analysis techniques that have been applied for BATSE observation of short duration 
511 keV transients (by David M. Smith) were applied.
``A posteriori" analyses of the background data at the explosion epoch of all 
the classical novae discovered optically during the whole period of CGRO 
operation (1991-2000), 
searching for some signal, were performed \citep{Her00}. Intervals of 12 hours were
analyzed at 6-hour spacing, to make sure that the peak of the outburst was not
splited. Background data were taken 24 hours before the period of interest.

The 3-$\sigma$ upper limits to the fluxes for the 3 most
promising novae of the whole sample, i.e. Nova Cyg 1992, Nova
Sco 1992 and Nova Vel 1999, were evaluated \citep{Her00}. 
All upper limits obtained were compatible with theory. 
The 3-$\sigma$ sensitivity using the 511 keV data
only is similar to that of \cite{Har99} with WIND/TGRS, but the sensitivity
of \cite{Har99} requires a particular line blueshift, whereas ours is
independent of it. 
It is important to mention that (as with TGRS in \cite{Har99})
detectability distances in \cite{Her00} were computed with the old 
$\gamma$-ray spectra models \citep{Gom98} previous to new determinations of 
$^{18}$F+p reaction rates, which reduced the $^{18}$F 
yields by a factor of $\sim 10$ \citep{Her99,Coc00}. So predicted fluxes should be reduced by 
a factor of $\sim 10$ and detectability distances by a factor of 
$\sim \sqrt{10}$ or even larger. 
$\gamma$-rays
Some preliminary results with RHESSI, which has 9 large, 
high-resolution, unshielded Ge detectors, were presented in the HEAD meeting in 2006 (Mathews, Smith, Hernanz; poster). 

A detailed study has been performed more recently with data from the Burst Alert Telescope (BAT) onboard the Swift satellite 
\citep{Sen08}. Because of its huge field of view, good sensitivity, and well-adapted (14-200 keV) energy band, Swift/BAT offers a new
opportunity for such searches. BAT data can be retrospectively used to search for prompt $\gamma$-ray emission from the direction of
novae after their optical discovery. The search for emission in the directions of the 24 classical novae discovered since the Swift launch 
yielded no positive results, but this was what could be expected, because none of these was close enough.

\section{Discussion. Current and future missions in the MeV range}
The long awaited detection of $\gamma$-rays from novae finally came with the Fermi/LAT; the emission observed 
corresponds to VHE $\gamma$-rays, produced either by neutral pions or by inverse Compton, as a consequence of particle acceleration 
in the shock wave ensuing the interaction between the nova ejecta and the red giant wind. RS Oph was predicted to emit such VHE photons 
\citep{TH07}, and V407 Cyg was indeed detected \citep{Abdo10}. A couple of new cases have just been reported in ATels (Nova Sco 2012 
and Nova Mon 2012). 

However, the detection of $\gamma$-rays in the MeV range, from nova radioactivities, has not been achieved yet. The predictions for the current INTEGRAL/SPI 
instrument are not very optimistic: distances shorter than 0.5 kpc for the $^7$Be line at 478 keV, and shorter than 1 kpc for the  $^{22}$Na line at 1275 keV, 
are required. A future generation of 
instruments is needed, either a powerful Compton Telescope (e.g., the ACT project) or a $\gamma$-ray lens - Gamma-Ray Imager, GRI - or a 
combination of both, DUAL (see papers by Boggs, von Ballmoos, and others, in ``New Astronomy Reviews", vol. 48, 2004 and in ``Experimental Astronomy", 
vol. 20, 2005). The Laue $\gamma$-ray lens provides up to now the best perspectives for detecting lines in the MeV range, 
since a large collecting area (the diffracting crystals acting as a photon concentrator) is combined with a small detector in its focal plane, thus yielding a good signal to noise ratio, not easily reachable with Compton Telescopes. We should wait several years until this already proven concept is feasible for a space mission.

\section*{Acknowledgements}
This research has been funded by the Spanish MINECO project AYA2011-24704, 
and the AGAUR (Generalitat of Catalonia) project 2009 SGR 315.

\label{lastpage}
\end{document}